# Tuning the exchange bias on a single atom from 1 mT to 10 T

Kai Yang[1], William Paul[1], Fabian D. Natterer[1,2], Jose L. Lado[3], Yujeong Bae[1,4,5], Philip Willke[1,4,5], Taeyoung Choi[4,5], Alejandro Ferrón[6], Joaquín Fernández-Rossier[7,8], Andreas J. Heinrich[4,5,*], and Christopher P. Lutz[1,*]

[1]IBM Almaden Research Center, San Jose, CA 95120, USA
[2]Physik-Institut, University of Zurich, CH-8057 Zurich, Switzerland
[3]Institute for Theoretical Physics, ETH Zurich, 8093 Zurich, Switzerland
[4]Center for Quantum Nanoscience, Institute for Basic Science (IBS), Seoul 03760, Republic of Korea
[5]Department of Physics, Ewha Womans University, Seoul 03760, Republic of Korea
[6]Instituto de Modelado e Innovación Tecnológica (CONICET-UNNE), and Facultad de Ciencias Exactas, Naturales y Agrimensura, Universidad Nacional del Nordeste, Avenida Libertad 5400, W3404AAS Corrientes, Argentina
[7]QuantaLab, International Iberian Nanotechnology Laboratory (INL), Avenida Mestre José Veiga, 4715-310 Braga, Portugal
[8]Departamento de Física Aplicada, Universidad de Alicante, San Vicente del Raspeig 03690, Spain

* Corresponding authors: A.J.H. (heinrich.andreas@qns.science) and C.P.L. (cplutz@us.ibm.com)

**Shrinking spintronic devices to the nanoscale ultimately requires localized control of individual atomic magnetic moments. At these length scales, the exchange interaction plays important roles, such as in the stabilization of spin-quantization axes, the production of spin frustration, and creation of magnetic ordering. Here, we demonstrate the precise control of the exchange bias experienced by a single atom on a surface, covering an energy range of four orders of magnitude. The exchange interaction is continuously tunable from milli-eV to micro-eV by adjusting the separation between a spin-1/2 atom on a surface and the magnetic tip of a scanning tunneling microscope (STM). We seamlessly combine inelastic electron tunneling spectroscopy (IETS) and electron spin resonance (ESR) to map out the different energy scales. This control of exchange bias over a wide span of energies provides versatile control of spin states, with applications ranging from precise tuning of quantum state properties, to strong exchange bias for local spin doping. In addition we show that a time-varying exchange interaction generates a localized AC magnetic field that resonantly drives the surface spin. The static and dynamic control of the exchange interaction at the atomic-scale provides a new tool to tune the quantum states of coupled-spin systems.**

Exchange interaction between magnetic atoms gives rise to exotic forms of quantum magnetism such as quantum spin liquids [1], and spin transport in magnetic insulators [2, 3]. It is also of great technological importance in tailoring magnetic devices [4-8]. For instance, in magnetic read heads, a ferromagnetic layer is "biased" to a specific magnetization direction by strong exchange coupling to an antiferromagnetic layer [4]. Weaker exchange interaction also plays an important role, in the spin dynamics of quantum magnets [9], in giant magnetoresistance devices [10], and in magnetic phases of coupled spins that depend on next-nearest-neighbor interactions such as spin chains [11] and spin glasses [12].

The size of the active center of electronic devices is moving toward the world of single atoms and single molecules, where magnetic nanostructures such as atomic dimers and clusters are contenders for novel data storage [13], spintronic devices [14, 15] and quantum computing applications [16]. When addressing the spin states of single atoms, the stability and orientation of the spin-quantization axis is critical [17]. The magneto-crystalline anisotropy usually defines the quantization axis in atoms [18, 19], but is sensitive to the electrostatic perturbations within the local crystal field [20, 21]. A versatile alternative to establish a preferred spin axis is to apply the exchange bias at the single-atom level [22, 23]. The exchange interaction, stemming from the overlap of electronic wave functions, is exponentially localized at the atomic scale and may thus be controlled over a large energy range by adjusting the interatomic distance [24, 25], providing a route towards tailored spin-based devices and materials [26].

Scanning tunneling microscopy (STM) is a powerful tool to study exchange interactions between atoms on surfaces [23, 27-29], by measuring changes in Kondo screening [27], energy relaxation times [23], and spin excitations using inelastic electron tunneling spectroscopy (IETS) [30]. To overcome the limitation of a discrete set of interatomic separations imposed by the substrate lattice, one spin center can be transferred to the STM tip, permitting continuous variation of the exchange interaction with surface adatoms [23, 28]. However, in previous studies, the precise characterization of the exchange interaction is indirect, and obscured by the competition with other interactions, such as the magneto-crystalline anisotropy present in large-spin systems [23], and the Kondo effect from the scattering electrons [27, 28].

Here, we choose a spin-1/2 atom, which is free of magneto-crystalline anisotropy [31], and decouple it from the metal substrate by using a thin insulator. This allows us for the first time to directly sense its exchange interaction with the spin center attached to the STM tip, by observing the position of the conductance steps in IETS. We observe an exponential decay of the exchange magnetic field as the tip is withdrawn from near point-contact with the atom.

However, the short decay length of the exchange interaction restricts IETS investigations to closely coupled spins, given its limited energy resolution (~meV) [32]. We extend our spectroscopic energy range further by 3 more orders of magnitude by employing electron spin resonance (ESR) in STM, which yields an energy resolution down to ~100 neV [31, 33-38]. We further show that the time-varying exchange interaction can be used to resonantly control a single spin.

We measure the coupling between a Ti and an Fe atom, where the Ti atom is placed on a bilayer MgO grown on Ag(001) and the Fe atom is attached to the metallic STM tip [Fig. 1(b)]. The unique advantage of this arrangement is the continuous variation of the interaction strength through the tip-sample distance $z$ [23, 28]. We study Ti atoms at the oxygen sites of the MgO lattice [31, 38]. The MgO layer improves the energy relaxation time ($T_1$) [39], which leads to sharper energy levels, such that spin transitions become accessible via IETS measurement at 0.6 K.

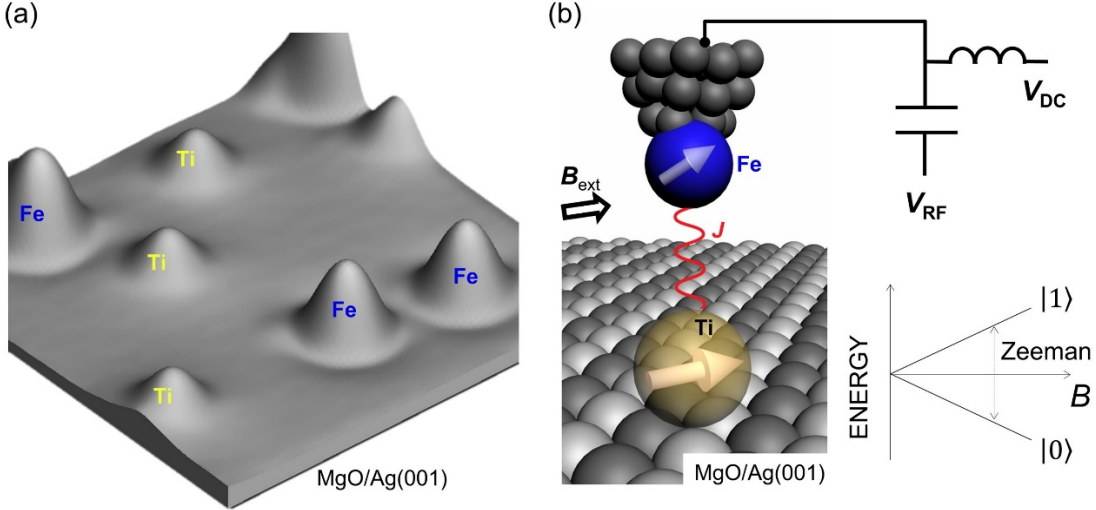

FIG. 1. Measurement setup of the exchange interaction between two atoms. (a) Constant current STM image of Ti and Fe atoms on a bilayer MgO on Ag(001) (set point: $V_{DC}$ = 50 mV, $I_{DC}$ = 10 pA). (b) Schematic showing a Ti atom on MgO and an Fe atom attached to the apex of the STM tip. The solid arrows indicate the orientations of the respective magnetic moments. The exchange interaction $J$ is indicated by the red curve. A radio-frequency voltage drives the ESR of the Ti atom. The external magnetic field $\mathbf{B}_{ext}$ is applied ~8° out of the plane of the substrate. Bottom right: Energy level diagram showing the Zeeman energy of the Ti spin as a function of magnetic field.

Each Ti adatom is a spin $S$ = 1/2 system (due to an attached hydrogen atom) [31] for which the absence of magneto-crystalline anisotropy [19, 31] means that the orientation of the quantization axis is determined by the effective magnetic field at the location of the Ti adatom. The magnetic field lifts the degeneracy of the two spin directions by the Zeeman effect, to give states labeled $|0\rangle$ and $|1\rangle$ [Fig. 1(b), lower right] having magnetic quantum numbers $m_s = -1/2$ and $+1/2$. In contrast, the Fe spin on the STM tip can be treated classically as a statistical average $\langle \mathbf{S}_{tip} \rangle$ [23, 40] due to the fast fluctuations of Fe spin, which results from the interaction with the conduction electrons in the metal tip. Due to the local magnetic anisotropy of the Fe atom, the direction of $\langle \mathbf{S}_{tip} \rangle$ is in general tilted from $\mathbf{B}_{ext}$. As is shown below, this tilting is essential to drive the ESR of the Ti.

We described the exchange coupling between Ti and the tip spin with $J \langle \mathbf{S}_{tip} \rangle \cdot \mathbf{S}$, where the coupling constant $J$ sensitively depends on the tip-Ti distance $z$. From the point of view of the Ti atom, this exchange coupling with the tip can be viewed as an effective magnetic field $\mathbf{B}_{tip} = J \langle \mathbf{S}_{tip} \rangle / (g \mu_B)$, where $g$ is the g-factor of the Ti spin, and $\mu_B$ is one Bohr magneton. The total magnetic field on the Ti atom is then $\mathbf{B} = \mathbf{B}_{\text{ext}} + \mathbf{B}_{\text{tip}}$. Here, the local magnetic field $\mathbf{B}_{\text{tip}}$ acts as an exchange bias on the Ti atom by modifying the Zeeman energy. We control $\mathbf{B}_{\text{tip}}$ by varying the tip height and characterize $\mathbf{B}_{\text{tip}}$ via IETS and ESR in the following.

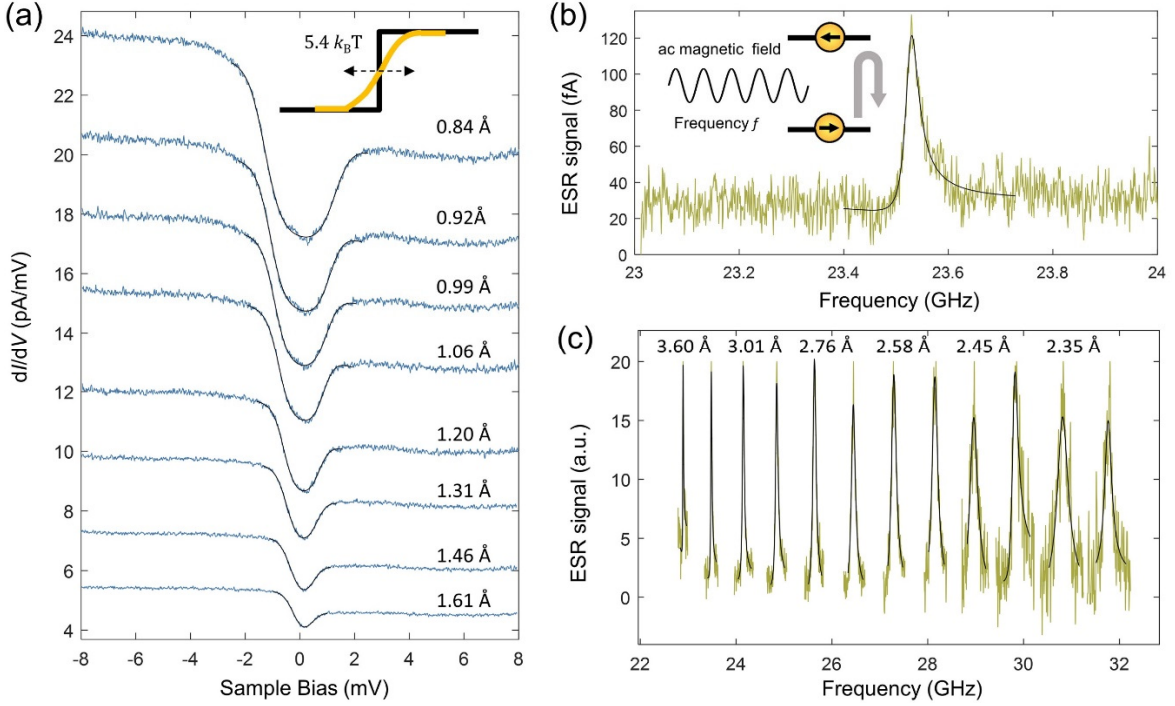

FIG. 2. IETS and ESR spectra as a function of tip-Ti distance. (a) IETS spectra on a Ti adatom using a magnetic tip ($T$ = 0.6 K). Black curves are fits to IETS line shapes [32]. Height $z$ = 0 corresponds to the point-contact (conductance of ~0.1 μS), and the decay constant of the junction conductance is ~0.52 Å [Fig. S1 [41]]. Inset: Schematic showing the broadening of the conductance step by the temperature. (b) ESR spectrum recorded on a Ti atom (set point: $V_{\text{DC}}$ = 50 mV, $I_{\text{DC}}$ = 10 pA, $V_{\text{RF}}$ = 20 mV, $T$ = 1.2 K). (c) ESR spectra at different tip heights (set point: $V_{\text{DC}}$ = 50 mV; $T$ = 0.6 K, $B_{\text{ext}}$ =0.9 T). The spectra are normalized to the same ESR amplitude for clarity. The black curves are fits to asymmetric Lorentzian line shapes [31].

IETS spectra on the Ti atom show a pair of conductance steps placed symmetrically about zero bias [Fig. 2(a), top], which originates from spin-flip excitations from $|0\rangle$ to $|1\rangle$. The steps are absent at zero magnetic field and shift to larger voltages at higher magnetic field, as was confirmed with IETS using a non-magnetic tip [42]. The IETS step position is a direct measure of the Zeeman splitting [42]. The asymmetry of step heights for opposite voltage polarities is due to the selection rule for spin excitations [43]. The ratio of the step heights yields the magnitude and sign of the spin polarization of the magnetic tip [43], which we find to be typically 15–35% for Fe-terminated tips.

As shown in Fig. 2(a), when the magnetic tip is moved farther away from the Ti atom (from z ≈ 0.84 Å to 1.61 Å), the conductance steps shift to lower energies until they become undetectable at ~0.2 meV, where the step width exceeds their separation. This reduction is a consequence of the decreasing magnitude of the tip magnetic field $B_{\text{tip}}$ as the tip is retracted. The step position is determined reliably when it is at an energy larger than the thermal line broadening $5.4k_{\text{B}}T \approx 0.28$ meV (at 0.6 K) [32]. When the tip is retracted farther, the determination of the step positions by IETS becomes difficult due to the thermally broadened Fermi-Dirac distributions of tip and sample [see error bars in Fig. 3(a)].

We continue to measure the Ti-tip exchange coupling for larger distances via the ESR technique [33], by sweeping the frequency of a radio-frequency (RF) voltage $V_{\text{RF}}$ [Fig. 1(b)]. Our use of ESR overcomes the temperature-limited energy resolution of IETS since the ESR resolution is determined by $T_2$, the quantum coherence time [33]. Here $T_2$ of the Ti spin is limited by the scattering of electrons that pass through the MgO barrier from the Ag substrate [39], and in this experiment, the energy resolution is much better than $k_{\text{B}}T$, by a factor of ~1000. The bandwidth of our ESR setup is limited to ~30 GHz, corresponding to ~0.12 meV as the highest ESR accessible energy. We next focus on the frequency of the ESR peak, which corresponds to the total Zeeman energy of the Ti spin. With larger tip-sample separation the Zeeman energy becomes inaccessible to the IETS measurements; and when the Zeeman splitting drops below ~0.12 meV it becomes accessible to our ESR investigation. In Fig. 2(c), we show that as the tip is retracted from $z \approx 2.3$ Å to 3.6 Å, the ESR peaks shift down from ~30 to 23 GHz (124 to 95 µeV).

Both the IETS step energies and the ESR peak positions directly indicate the total Zeeman energy ($E_{\text{total}}$) experienced by the Ti atom, including the Zeeman energy due to both $\mathbf{B}_{\text{ext}}$ and $\mathbf{B}_{\text{tip}}$: $E_{\text{ext}} = g\mu_{\text{B}}B_{\text{ext}}$ and $E_{\text{tip}}(z) = g\mu_{\text{B}}B_{\text{tip}}(z)$. Assuming exchange coupling between the tip and the Ti atom, we expect exponential dependence on the tip height: $E_{\text{tip}}(z) \propto \exp(-z/d_{\text{exch}})$. In Fig. 3(a), we display $E_{\text{total}}$ as a function of tip-Ti distance measured by IETS and ESR. By fitting $E_{\text{total}}$ with an exponential function and a vertical offset, we find the decay length $d_{\text{exch}} = 0.42 \pm 0.02$ Å. Different magnetic tips show similar decay constants [Fig. S3 [41]]. The asymptotic value of $E_{\text{total}}(z)$ (~92.9 µeV) corresponds to the absence of the tip field and represents $E_{\text{ext}}$ alone [Fig. 3(a)]. This allows us to calculate the g-factor of the Ti spin, which yields $g = E_{\text{ext}}/(B_{\text{ext}}\mu_{\text{B}}) = 1.8$, in agreement with the independent measurement of dipole-coupled Ti-Ti atoms on MgO [31]. Note that the monotonic increase of $E_{\text{total}}$ with decreasing tip height indicates ferromagnetic coupling between the tip and Ti spin, which probably arises from the direct exchange interaction.

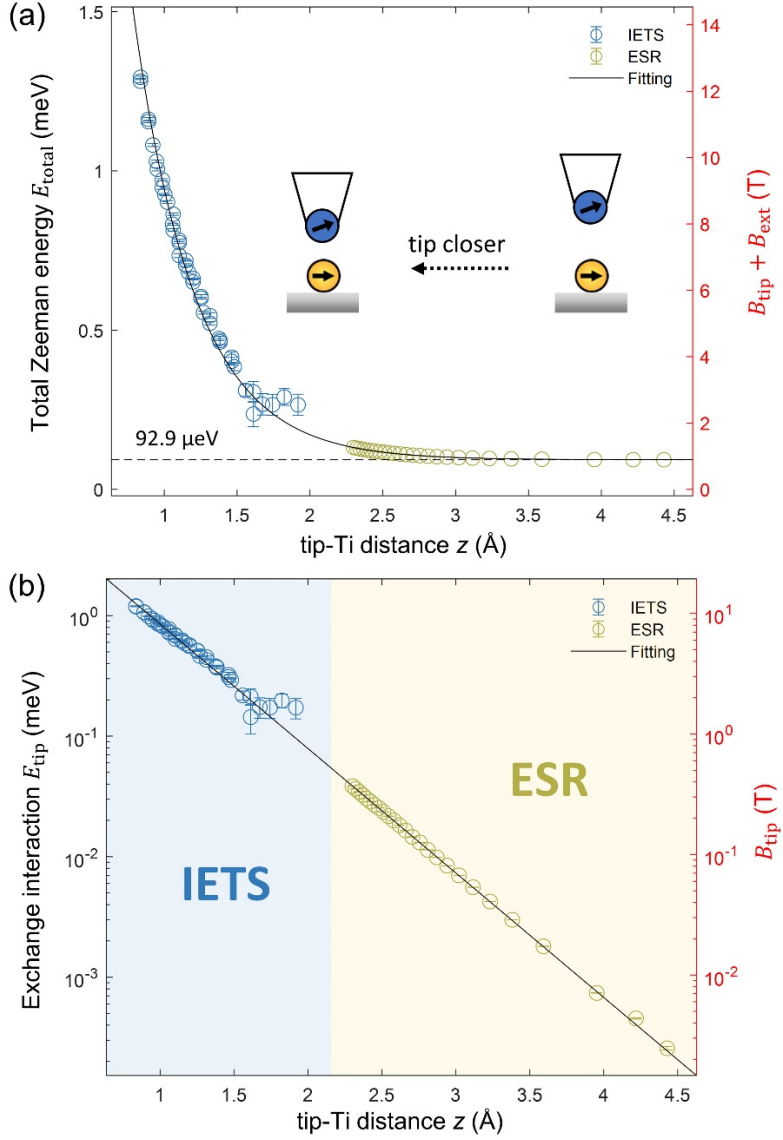

FIG. 3. Exchange interaction as a function of tip-to-atom distance. (a) $E_{\text{total}}$ of a single Ti atom under a magnetic tip as a function of the tip-Ti distance. Symbols represent the Zeeman energy determined from IETS (blue circles) and ESR (light-green circles). The solid line is an exponential fit. The asymptotic value is 92.9 $\mu$eV indicated by the dashed line. (b) $E_{\text{tip}}$ and $B_{\text{tip}}$ as a function of the tip-Ti distance (log scale on the $y$ axes). The data are extracted from (a) by subtracting the asymptotic value. The solid line is an exponential fit. The error bars are determined by the fitting uncertainties of the IETS step positions and the ESR frequencies.

We extract the exchange coupling energy $E_{\text{tip}}(z)$ from $E_{\text{total}}(z)$ by subtracting the asymptotic value $E_{\text{ext}}$, and plot it in Fig. 3(b). The exchange coupling notably covers the energy range of four orders of magnitude from $\sim 10^{-7}$ eV to $\sim 10^{-3}$ eV [Fig. 3(b), left axis]. The effective tip field $B_{\text{tip}}$ is then calculated as $B_{\text{tip}}(z) = E_{\text{tip}}(z)/(g\mu_B)$, giving a range from $\sim 1$ mT to $\sim 10$ T [Fig. 3(b), right axis]. This tip magnetic field with a large dynamic range could thus be used to stabilize single atom spins in weak or zero externally applied fields. It also provides a large tunability of quantum states of coupled atoms by controlling the Zeeman energy of an individually selected atoms [23, 31]. Sweeping the tip field by moving the tip can also be used to achieve ESR, by tuning the Zeeman energy into resonance with a fixed-frequency RF voltage, which provides some technical advantages over sweeping the frequency of the RF voltage [44].

Note that fitting $E_{\text{total}}(z)$ by considering the tilting of $\mathbf{B}_{\text{tip}}$ with respect to $\mathbf{B}_{\text{ext}}$ yields similar decay constant $d_{\text{exch}}$ and a tilting angle $\varphi$ of ~60 degrees for this tip [Fig. S2 [41]]. Including magnetic dipolar coupling in the model improves the fitting only marginally, for distances larger than 4 Å. Thus, we didn't consider the dipolar coupling in the fitting in Fig. 3.

In addition to the static control of the Zeeman energy by the tip exchange field, a time-varying exchange field allows resonant control the Ti spin states, which makes the ESR measurement possible. The AC electric field due to $V_{\text{RF}}$ drives ESR, but does not couple to the spin directly. Instead, to drive single-spin transitions, the AC electric field produces an effective AC magnetic field, $B_{\text{AC}}(t) = B_{\text{AC}} \cos(2\pi f t)$, oscillating at the Larmor frequency in the plane perpendicular to the quantization axis of the spin. On resonance, the Rabi frequency $\Omega$ is proportional to $B_{\text{AC}}$. For Ti on MgO, the measured $\Omega$ at the tip-Ti distance $z$ = 4.3 Å is ~2.8 rad/μs (at $V_{\text{RF}}$ = 10 mV; see Fig. S3 [41]), corresponding to an effective AC magnetic field of ~0.03 mT. The origin of this field cannot be the oscillating tunneling current (<1 pA) or the displacement current (~1 pA) induced by the RF voltage. The resulting magnetic field amounts to only ~$10^{-6}$ mT (assuming an atomic radius of 1 Å), as calculated by Maxwell's equations (Supplemental Sec. 8 [41]).

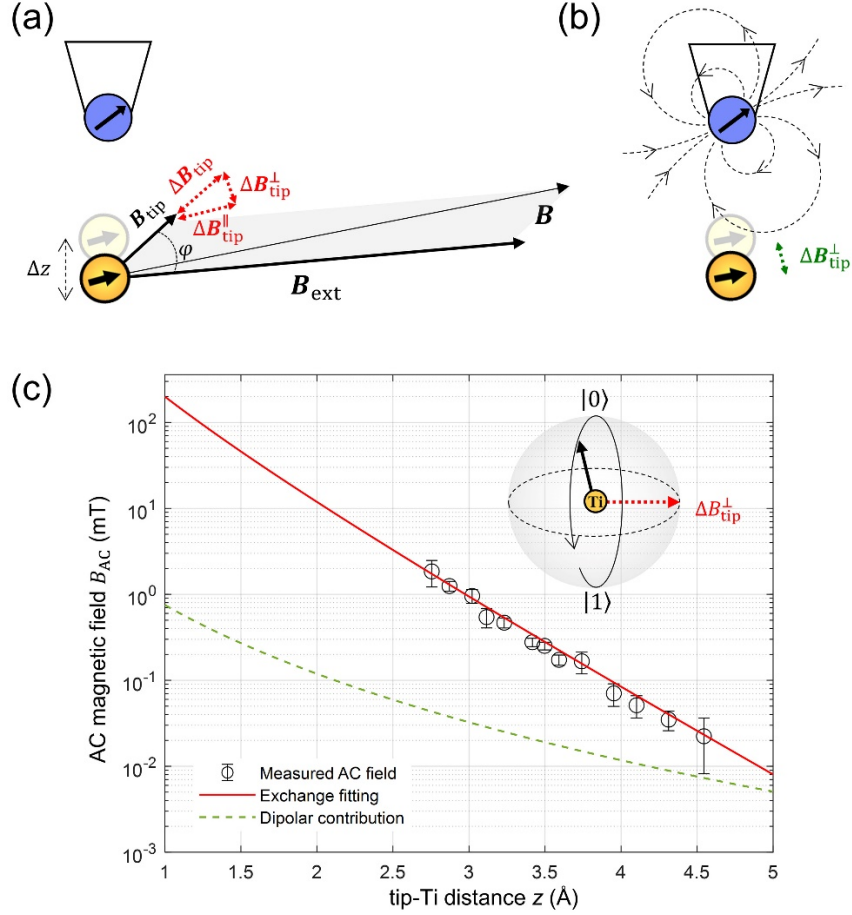

FIG. 4. Driving ESR by varying tip-Ti magnetic interactions. (a) and (b), Schematic of two possible ESR driving mechanisms: modulation of the (a) exchange and (b) magnetic dipolar interactions. (c) Extracted AC magnetic field as a function of tip-Ti distance $z$ at $V_{RF}$ = 10 mV. Red solid curve is the fit considering exchange interactions (with $\varphi \approx 60°$). The green dashed curve is the calculated AC magnetic field from dipolar interaction. Inset shows the rotation of the Ti spin (solid black arrow) around the AC tip field (dashed red arrow) between $|0\rangle$ and $|1\rangle$ states in a Bloch sphere in the rotating frame.

We propose that the ESR transitions are driven by the modulation of $\mathbf{B}_{tip}$ [40]. The AC electric field induces a small z-axis displacement of the Ti atom [40, 45]. Since $\mathbf{B}_{tip}$ is strongly spatially inhomogeneous, the vertical displacement $\Delta z$ results in a time-varying magnetic field having a component that is perpendicular to the total field $\mathbf{B}$ [Fig. 4(a)]. In the following, we use the effective driving AC magnetic field derived from the measured Rabi rate, and then extract the corresponding $\Delta z$ by using the measured $z$-dependent exchange interaction.

The spin Hamiltonian of Ti influenced by an RF voltage at frequency $f$ is:

$$H = g\mu_B \mathbf{B} \cdot \mathbf{S} + g\mu_B \Delta \mathbf{B}_{tip} \cdot \mathbf{S} \cos(2\pi f t) \qquad (1)$$

The first term determines the quantization axis of the Ti spin, which is along the direction of the total magnetic field $\mathbf{B} = \mathbf{B}_{\text{ext}} + \mathbf{B}_{\text{tip}}$ [Fig. 4(a)]. Here $\mathbf{B}_{\text{tip}}$ is tilted with respect to $\mathbf{B}$, by an angle given approximately by $\varphi$ since $B_{\text{tip}} \ll B_{\text{ext}}$ when performing ESR spectra. The oscillating tip field $\Delta \mathbf{B}_{\text{tip}} \cos(2\pi f t)$ has a component $\Delta B_{\text{tip}}^{\perp} \cos(2\pi f t)$ that is perpendicular to $\mathbf{B}$, which corresponds to $B_{\text{AC}}(t)$ that drives the spin resonance. The amplitude of the driving field $\Delta B_{\text{tip}}^{\perp}$ is related to the zero-to-peak displacement of the Ti atom ($\Delta z$) at tip-Ti distance ($z$) by

$$\Delta B_{\text{tip}}^{\perp}(z) = \frac{\partial B_{\text{tip}}}{\partial z} \Delta z \sin \varphi \approx -\frac{B_{\text{tip}}}{d_{\text{exch}}} \Delta z \sin \varphi \propto \frac{\exp(-z/d_{\text{exch}})}{z} \quad (2)$$

Here $\varphi \approx 60°$ as determined by fitting $E_{\text{total}}$ considering the tilting of $\mathbf{B}_{\text{tip}}$ (Supplemental Sec. 3 [41]). $d_{\text{exch}}$ is the decay length of the exchange interaction between the tip and the Ti atom, which is ~0.42 Å, obtained in Fig. 3. The vertical displacement varies as $\Delta z \propto z^{-1}$, by assuming that $\Delta z$ results from the AC electric field $V_{\text{RF}}/z$. We see $\Delta B_{\text{tip}}^{\perp}(z)$ has the same exponential dependence on $z$ as the exchange coupling.

To obtain the value of the vertical displacement $\Delta z$, we fit the model for $\Delta B_{\text{tip}}^{\perp}(z)$ of Eq. (2) to the values of $B_{\text{AC}}$ determined experimentally [Fig. 4(c), red curve]. The values for $B_{\text{AC}}$ were extracted from the measured values of $\Omega$ by using $B_{\text{AC}} = 2\hbar\Omega/(g\mu_B)$ (Supplemental Sec. 5 [41]). As shown in Fig. 4(c), $\Delta B_{\text{tip}}^{\perp}(z)$ describes the trend well for $B_{\text{AC}}$ as the tip-Ti distance changes, and the fitting yields $\Delta z = 2.9 \pm 0.2$ pm at the tip-Ti distance of 4.3 Å for $V_{\text{RF}}$ = 10 mV.

If we assume that this displacement comes exclusively from the stretching of the Ti-O bond, we can infer a restoring force $k\Delta z = q_{\text{Ti}} V_{\text{RF}} \cdot z^{-1}$. Taking $q_{\text{Ti}} \approx 1e$, and using $k = m(2\pi\nu)^2$, where $m$ is the mass of the Ti atom, we obtain a stretching frequency $\nu \approx 1$ THz. We computed the Ti-O stretching frequency using density functional theory (DFT) calculations (Supplemental Sec. 7 [41]) and obtained $\nu \approx 4$ THz. This larger frequency requires a larger restoring force, and thereby a smaller displacement than in the experiment. This indicates that other sources of $\Delta z$, in addition to the Ti-O stretching, must contribute significantly. Specifically, the relative vertical displacement of the Ti atom with respect to the tip could be enhanced by the piezo electric motion of the MgO layer, as well as the motion of the Fe atom at the tip apex [46].

The AC magnetic field might also have a contribution from the magnetic dipolar interaction between the tip and the Ti atom [Fig. 4(b)], though at these sub-nanometer distances we generally expect exchange interaction to exceed dipolar interaction [31]. The magnetic dipolar field would contribute an AC magnetic field proportional to $(z + d_0)^{-4} \Delta z \sin \varphi$ ($d_0 \approx 2$ Å is the diameter of an atom) [Fig. S4 [41]]. We estimate this AC dipolar magnetic field by using the $\Delta z$ obtained above [Fig.

4(c), green curve]. The modulation of the tip exchange field gives a better description of the slope of $B_{AC}$ as tip-Ti distance changes. The tip exchange interaction is also more effective to drive ESR of Ti on the surface, though Fig. 4(c) shows that the dipolar magnetic field may contribute at the largest separations.

We have demonstrated that the local exchange bias on a single atom can be deliberately tuned with high precision and over many orders of magnitude. Our choice of a spin-1/2 system permits the direct visualization of the exchange interaction, by observing the energy of the spin-excitation steps and the frequency of the ESR peaks. Our quantitative analysis shows that the electric field drives ESR by modulating the exchange magnetic field, which should be applicable to spin resonance in other solid-state spin systems, such as molecular magnets [16] and quantum dots [47]. The static and dynamic control of the exchange interaction permits local probing and tuning of coupled quantum spins by applying a local magnetic field on a single atom [23, 31]. Use of a more precisely controlled magnetic tip (for example, a tip functionalized by attaching a magnetic molecule [48]), may allow the imaging of the 3D distribution of the spin-polarized atomic orbitals of a single atom on a surface.

We thank Bruce Melior for expert technical assistance. We gratefully acknowledge financial support from the Office of Naval Research. W.P. thanks the Natural Sciences and Engineering Research Council of Canada for fellowship support. F.D.N. appreciates support from the Swiss National Science Foundation under project numbers PZ00P2_167965 and PP00P2_176866. A.F. acknowledges CONICET (PIP11220150100327) and FONCyT (PICT-2012-2866). Y.B., P.W., T.C. and A.J.H acknowledge support from IBS-R027-D1. J.L.L. acknowledges support from the ETH Fellowship program. J.F-R. thanks FCT under the project "PTDC/FIS-NAN/4662/2014" as well as Generalitat Valenciana funding Prometeo2017/139 and MINECO Spain (Grant No. MAT2016-78625-C2).